\documentclass[letterpaper,twocolumn,english,aps,prl,groupedaddress,superscriptaddress,showpacs,amsfonts]{revtex4}
\usepackage{ae}
\usepackage{aecompl}
\usepackage[T1]{fontenc}
\usepackage[latin1]{inputenc}
\usepackage{float}
\usepackage{amsmath}
\usepackage{graphicx}
\usepackage{amssymb}

\makeatletter
\usepackage{babel}
\makeatother
\begin{document}

\title{A Tale of Two Theories: Quantum Griffiths Effects in Metallic Systems}

\author{A.~H.~Castro Neto}

\affiliation{Department of Physics, Boston University, 590 Commonwealth Ave.,
  Boston, MA 02215}

\author{B.~A.~Jones}

\affiliation{IBM Almaden Research Center, San Jose, CA 95120-6099}

\begin{abstract}
We show that two apparently contradictory theories on the 
existence of Griffiths-McCoy singularities in magnetic metallic 
systems \cite{prb,MMS2}  are in fact mathematically equivalent. 
We discuss the generic phase diagram of the problem and show
that there is a non-universal crossover temperature
range $T^* < T < \omega_0$ where power law behavior (Griffiths-McCoy
behavior) is
expect. For $T<T^*$ power law behavior ceases to exist
due to the destruction of quantum effects generated by the
 dissipation in the metallic environment. We show that $T^*$ is an analogue
of the Kondo temperature and is controlled by 
non-universal couplings. 
\end{abstract} 

\maketitle

\begin{figure}[htb]
\begin{center}
\includegraphics[width=8cm,keepaspectratio]{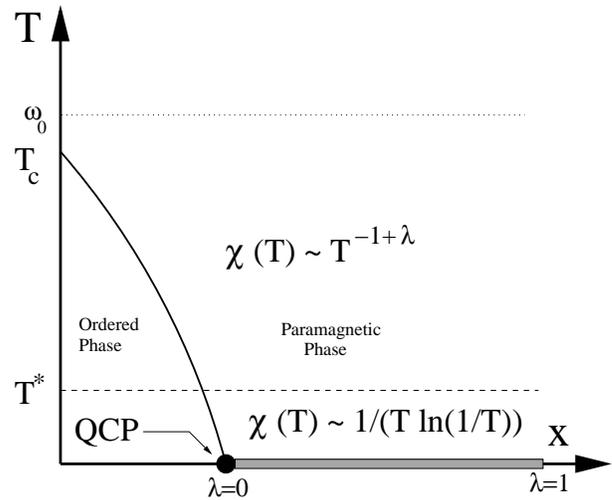}
\end{center}
\caption{Schematic phase diagram $T \times x$ where $x$
controls the phase transition. Also shown is the behavior
of the magnetic susceptibility, $\chi(T)$, in the proximity of the
quantum critical point.}
\label{phd}
\end{figure}

The problem of non-Fermi liquid behavior (NFL) in U and Ce intermetallics
continues to attract a lot of attention due to the 
breakdown of Landau's Fermi liquid theory in metallic alloys \cite{greg}.
NFL is often characterized by power law or logarithmic temperature 
behavior in physical quantities such as the magnetic susceptibility, $\chi(T)$
and specific heat, $C_V(T)$. Many NFL materials have as common features
the closeness to a magnetic phase transition and disorder generated
by the alloying \cite{doug}. It has been proposed
that NFL behavior can be associated in some materials
with quantum Griffiths-McCoy 
singularities close to a quantum critical point (QCP) \cite{prl}.
These singularities are related with the tunneling, at low temperatures, 
of magnetic clusters with $N$ spins generated by the percolating nature of the
magnetic phase transition. As a result, the physical properties acquire
non-universal power law behavior, 
$\chi(T) \propto C_V(T)/T \propto T^{-1+\lambda}$,
with the exponent $\lambda<1$ dependent on the distance from the QCP. 
While in its first 
version \cite{prl} the theory did not consider properly the dissipation 
coming from the electronic degrees of freedom, it was extended 
to include dissipation \cite{prb} with the final conclusion that power 
law behavior disappears below a crossover temperature $T^*$ due to
the freezing of the magnetic clusters  (that is, the quantum 
Griffiths singularities are suppressed close enough to the QCP). Instead 
of power law, it was predicted that for $T<T^*$ one should
have $\chi(T) \propto 1/(T \ln(1/T))$ 
and $C_V(T)/T  \propto 1/(T \ln^2(1/T))$ (see fig.\ref{phd}). 
These results seem to explain a stronger than
power law divergence and also the discrepancy between the specific heat and susceptibility
exponents at low temperatures in certain materials \cite{marcio}. 
Nevertheless, for $T^*<T<\omega_0$ it was found that 
$\chi(T) \propto T^{-1+\lambda}$, that is, quantum Griffiths behavior. 
Besides the temperature dependence the theory has also predicted 
a magnetic field, $H$, dependence of 
the magnetization, $M(H,T)$, and $C_V(H,T)$
with $M(H,T) \propto H T^{-1+\lambda}$
and  $C_V(H,T)/T \propto T^{-1+\lambda}$
for $T>(H,T^*)$ and $M(H,T) \propto H^{\lambda}$ 
and $C_V(H,T)/T \propto T^{-3+\lambda/2} H^{2+\lambda/2} e^{-H/T}$
for $H>T>T^*$. These 
predictions seem to be in agreement with the experimental data in a series of 
different materials where disorder plays a fundamental role 
\cite{greg,greg1}.

Using the Hertz theory of quantum critical phenomena \cite{hertz} 
and treating the problem of an Ising field theory in the presence of a 
single impurity, Millis, Morr, and Schmalian (MMS) \cite{MMS1} have cast 
doubts on the possibility of Griffiths singularities in these materials 
based on the over-damping of the cluster (droplet) dynamics due to the 
electronic degrees of freedom. This effect is the result of 
the slow decay of the droplet profile in the single impurity case. 
MMS have extended their calculation to a finite density of
impurities \cite{MMS2} and conclude that over-damping destroys quantum Griffiths
singularities leading to $\chi(T) \propto 1/T$ times $\ln(T)$ corrections
at all $T<\omega_0$ where $\omega_0$ is the high temperature cut-off of the theory
above which the droplets are thermally activated. This result is 
in agreement with 
predictions in ref.~\cite{prb} 
that $\chi(T) \propto 1/(T \ln(1/T))$ at low temperatures. 
MMS, however, do not observe
any crossover with $T$ (as it was found in ref.~\cite{prb}) 
and conclude that power law behavior should not be observed. 

In what follows we demonstrate the mathematical equivalence of 
the theory proposed in ref.~\cite{prb} and by MMS \cite{MMS2}.  
Both theories have 
a non-universal energy scale, $T^* =\omega_0 e^{-C_2} < \omega_0$, 
so that for $T^*<T<\omega_0$ one has $\chi(T) \propto T^{-1+\lambda}$ 
and therefore quantum Griffiths singularities. $T^*$ plays the role
of an effective Kondo temperature of the droplet that is a non-universal
quantity dependent on the constant $C_2$ defined below. As we show below,
$C_2$ depends on the microscopic couplings of the material and 
changes from system to system.
While MMS take $C_2 \approx 1$ \cite{MMS2} leading to a narrow
region of quantum Griffiths behavior,  
we have argued in ref.~\cite{prb} that $C_2 \gg 1$ leading
to $T^* \ll \omega_0$ and therefore to a sizeable region in 
$T$ where power law behavior should be observed. 
We believe this result to be in agreement with
experiments in various different materials\cite{greg,marcio}. 
In this paper we revisit the theory proposed by MMS and study the size of 
the Griffiths region (power law behavior) as a function of $C_2$. 
We will show that this region shrinks very fast with the decrease of
$C_2$.

The starting point is the Hertz action for a field theory with
Ising symmetry \cite{hertz}, $S=S_0+S_I$, where:
\begin{eqnarray}
S_0 &=& \frac{E_0 \xi_0^2 T}{8 \pi} 
\sum_{n,{\bf k}} \left(\xi^{-2} + k^2 + 
\frac{\omega_n^2}{c^2} 
+ \frac{8 \pi |\omega_n|}{\Gamma E_0^2}\right) |\phi({\bf k},\omega_n)|^2
\nonumber
\\
S_I &=& \frac{E_0}{16 \pi} \int_0^{\beta} d\tau \int d{\bf r} 
\, \, \, \phi^4({\bf r},\tau) \, ,
\label{basic}
\end{eqnarray}
here $E_0$ is a characteristic energy scale (assumed to be of the order 
of the Kondo temperature of the system), $\xi_0$ is a characteristic 
length scale (of the order of the lattice spacing), $\xi$ is the correlation 
length, $c$ is a characteristic velocity, $\Gamma$ is the damping coefficient,
and $\phi({\bf r},\tau)$ is the 
order parameter ($\phi({\bf k},\omega_n)$ its Fourier transform in momentum 
and Matsubara frequency). In principle the quantities that appear in 
the action have to be obtained from a microscopic theory or to be used
as fitting parameters to the experiments. 

The main difference between the two approaches discussed in ref.~\cite{prb} and ref.~\cite{MMS2} 
is that ref.~\cite{MMS2} describes a soft spin model in the continuum 
while in ref.~\cite{prb} a hard spin approach
in a lattice is used. In the first case the amplitude of the droplets, $\phi_0$,
can fluctuate statistically 
while in the second case the amplitude of a cluster of $N$ spins is fixed
by the spin $S$ of the spins in the cluster.
Using the single impurity solution discussed in ref.~\cite{MMS1} as a variational
ansatz, MMS derived the probability distribution, $P(R,\phi_0)$ for a droplet of 
size $R$ and amplitude $\phi_0$, and the renormalized droplet tunneling
splitting, $\omega_{{\rm tun}}$.
 
In order to make the connection between the different theories
we notice that the number of spins in a droplet is approximately given 
by: $\int d{\bf r} \phi({\bf r})/(4 \pi \xi_0^3/3) \propto 
\phi_0 (R/\xi_0)^3$, 
where it is assumed that the droplet exists in a region of size $R$ only. 
In what follows, instead of using MMS's reduced variables: 
$y=R/\xi$, $f = \phi_0 (\xi/\xi_0)$, and $u=V_0^2 (\xi/\xi_0)$ ($V_0$ 
is the strength of the disorder) we define a new variable: 
\begin{eqnarray}
N(y,f) = f^2 \frac{\xi}{\xi_0} y^3 \, ,
\label{nyf}
\end{eqnarray}
and new parameters:
\begin{eqnarray}
\nu &=& c_{\gamma} a C_2 \, ,
\nonumber
\\
N_c &=& (c_{\gamma} a)^{-1} \, ,
\label{cv}
\end{eqnarray}
where
\begin{eqnarray}
a(y) &=& 1+3/y^2 \, ,
\nonumber
\\
c_{\gamma} &=& E_0/(6 \Gamma) \, ,
\nonumber
\\
C_2 &=& 1 + \ln\left[E^2_0 \xi_0^2/(6 c^2)\right] \, .
\label{c2}
\end{eqnarray}
Notice that the parameters defined in (\ref{cv}) are
independent of $R$ and $\phi_0$. Using these variables
we can rewrite the main results of ref.~\cite{MMS2}, namely,
the droplet tunneling splitting (eq.~(36) of ref.~\cite{MMS2}),
\begin{eqnarray} 
\omega_{{\rm tun}}(N) = \omega_0 \exp\left\{-\frac{\nu N}{1-N/N_c}\right\} \, ,
\label{deltan}
\end{eqnarray}
and the probability distribution 
($N[y^3,f^2]$ in eq.~(15) of ref.~\cite{MMS2}), 
\begin{eqnarray}
P(N,f) \propto N^{1-\theta} e^{-N/N_{\xi}(f)}
\label{pn}
\end{eqnarray}
where $\theta = 5/2$ and 
\begin{eqnarray}
N_{\xi}(f) = \frac{f^2 u}{(f^2+a)^2} \frac{\xi}{\xi_0} \, . 
\label{nxi}
\end{eqnarray} 

We can now make a direct comparison between the two different
theories. Notice that the tunneling splitting is only
dependent on the number of spins in the droplet, $N$, and
it is identical to eq.~(5.6) of ref.~\cite{prb} if we
assume a single cut-off energy scale (that is, set $W=\omega_0$
and $\varphi=1$, see eq.~(4.84) in ref.~\cite{prb}). This result
allows us to identify $N_c$ as the maximum number of spins
in the droplet such that, above this number, tunneling ceases
to occur, and $\nu$ is the effective damping coefficient (the
equivalent of $\gamma$ in eq.~(5.6) of ref.~\cite{prb}). 
The probability distribution (\ref{pn}) is the equivalent
of eq.~(5.7) of ref.~\cite{prb} except that in ref.~\cite{MMS2}
this probability distribution 
depends not only on the number of spins in the droplet
but also on their amplitude. Although in a soft spin model
there are fluctuations in the size of the droplet, those
fluctuations are limited by the exponential term in (\ref{pn}).
In fact, MMS find that the droplets that contribute most
to the magnetic susceptibility are such that 
\begin{eqnarray}
f_{{\rm MMS}} \propto \left(\frac{\xi_0}{\xi}\right)^{1/2} \, .
\label{fmms}
\end{eqnarray}
If one replaces (\ref{fmms})
into eq.~(\ref{nxi}) we find, for $\xi \gg \xi_0$,
\begin{eqnarray}
N_{\xi,{\rm MMS}} \propto V_0^2 \frac{\xi}{\xi_0} \, .
\label{nximms}
\end{eqnarray}
In ref.~\cite{prb} it is shown that (see eq.~(4.6) of ref.~\cite{prb}):
\begin{eqnarray}
N_{\xi,{\rm CNJ}} \propto \left(\frac{\xi}{\xi_0}\right)^{D} \, ,
\label{ncnj}
\end{eqnarray}
where $D=2.54$ is the fractal dimension of the cluster in three
dimensions. Therefore, (\ref{nximms}) and (\ref{ncnj}) are very similar 
(the difference in the exponents is due to the different
nature of the percolation in the two problems) and
show that $N_{\xi}$ diverges as $\xi \to \infty$. 

It should be clear at this point
that the two main elements of the theory, namely the
tunneling splitting and its distribution, although calculated in
very different way, are essentially the same in both theories. 
The main question is why MMS reach the conclusion that power law
behavior, that is, quantum Griffiths singularities cannot be
observed while in ref.~\cite{prb} it was found that there is a
wide range in $T$ where power law behavior should be observed. 
To answer this question let us consider the distribution of
tunneling splittings, that is, the probability of finding a droplet
with tunneling splitting between $\Delta$ and $\Delta+d\Delta$, 
(instead of the distribution of cluster sizes).
It is easy to convert from one to the other using (\ref{deltan})
and (\ref{pn}) (we define $\Delta=\omega_{{\rm tun}}$):
\begin{eqnarray}
P(\Delta) &=& P(N(\Delta)) \left|\frac{d N}{d \Delta}\right|
\nonumber
\\
&\propto& \frac{(\ln(\omega_0/\Delta))^{1-\theta} 
\exp\left\{-\frac{N_c}{N_{\xi}}
\frac{\ln(\omega_0/\Delta)}{N_c \nu + \ln(\omega_0/\Delta)}\right\}
}{\Delta \left(N_c \nu + \ln(\omega_0/\Delta)\right)^{2-\theta}} \, .
\label{maindish}
\end{eqnarray}
It is straightforward to see that there is a characteristic tunneling splitting $\Delta_c$,
\begin{eqnarray}
\Delta_c = \omega_0 e^{-\nu N_c} \, ,
\label{deltac}
\end{eqnarray}
such that for $\Delta \gg \Delta_c$ one has:
\begin{eqnarray}
P(\Delta) \propto \Delta^{1/(\nu N_{\xi})-1} \, ,
\end{eqnarray}
which leads to $\chi(T) \propto T^{-1+\lambda}$ for $T>\Delta_c$ 
in agreement with eq.~(6.1) of 
ref.~\cite{prb}. Using (\ref{cv}) and (\ref{nximms})
we find:
\begin{eqnarray}
\lambda_{{\rm MMS}} = \frac{1}{\nu N_{\xi}} \propto \frac{\xi_0}{V_0^2 \xi} \, ,
\end{eqnarray}
which is eq.~(44) of ref.~\cite{MMS2}. This last result can be 
compared with eq.~(6.1) of ref.~\cite{prb} where:
\begin{eqnarray}
\lambda_{{\rm CNJ}} \propto \left(\frac{\xi_0}{\xi}\right)^{D} \, ,
\end{eqnarray}
and both theories predict that $\lambda \to 0$ as one
approaches the QCP (see fig.\ref{phd}).
This behavior is characteristic of quantum Griffiths
singularities. Once again the difference in the dependence of $\lambda$
with the dimensionality comes from the fact that ref.~\cite{prb} deals with
a percolation problem in a lattice while ref.~\cite{MMS2} studies an
impurity problem in the continuum. Nevertheless, the overall conclusion
that $\lambda$ vanishes at the critical point is common to both theories.
On the other hand, if $\Delta \ll \Delta_c$ we have from (\ref{maindish}):
\begin{eqnarray}
P(\Delta) \propto \frac{1}{\Delta \ln(\omega_0/\Delta)}
\end{eqnarray}
which leads to the result that $\chi(T) \propto 1/(T \ln(1/T))$
in agreement with eq.~(5.18) in ref.~\cite{prb} and with
the conclusion expressed in ref.~\cite{MMS2} that the susceptibility is $1/T$
times logarithms. Thus, we have unequivocally shown that the theory
presented by MMS has a crossover temperature $T^*=\Delta_c$ from power
law to $1/(T \ln(1/T))$ from high to low temperatures
and also proved the equivalence of the results of ref.~\cite{prb}
and ref.~\cite{MMS2}.

The reason why the crossover is not discussed in ref.~\cite{MMS2}
can be easily understood if we rewrite $\Delta_c$ in (\ref{deltac})
in terms of the parameters in (\ref{cv}):
\begin{eqnarray}
\Delta_c = \omega_0 \, \, e^{-C_2}
\end{eqnarray}
which means that the crossover depends on the value of $C_2$, that is, 
depends on a non-universal number. By assuming that $c/\xi_0 \approx E_0$
MMS find $C_2 \approx 1$. This choice 
implies that $\Delta_c/\omega_0 \approx 0.4$ which
is not particularly small and leads to a limited dynamical
range for the power law to be observed (in Fig.3 of ref.\cite{MMS2} 
the plots stop at frequencies of order of $0.1 \omega_0$ while
the crossover should be observed at $0.4 \omega_0$). 
Thus, it is of great interest
to understand how the magnetic response of the system depends on
$C_2$. 

\begin{figure}[htb]
\begin{center}
\includegraphics[width=6cm,keepaspectratio,angle=-90]{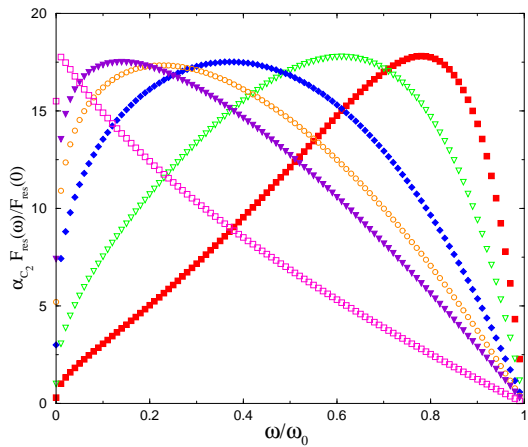}
\end{center}
\caption{Plot of $J_{{\rm res}}(\omega)/J_{{\rm res}}(0)$ as a
function of $\omega/\omega_0$ for $C_2=10$ (open squares),
$C_2=4$ (closed triangles), $C_2=3$ (open circles), $C_2=2$ (closed diamonds),
$C_2=1$ (open triangles), $C_2=0.1$ (closed squares). The other
parameters are given in the text. The crossover frequency
from over-damping to under-damping is roughly 
the frequency associated with the maximum in
each curve.}
\label{ires}
\end{figure}

We see from (\ref{c2})
that $C_2$ depends on the details of the lattice through $\xi_0$ and $c$ as
well as the Kondo energy scale through $E_0$. 
Any physical crossover, by definition, 
depends on non-universal parameters. A
famous example is the Kondo problem where the Kondo temperature, 
$T_K = \omega_0 e^{-1/g}$ (where $g$ is the strength of the
spin exchange interaction between localized moments and conduction electrons),
is also a non-universal function of the cut-off and, like the problem
at hand, is an exponential function of effective coupling.
As in the Kondo case the crossover temperature in this problem is 
exponentially sensitive on the choice of $C_2$. The crossover
can be easily visualized if one considers, for instance, the
function $J_{{\rm res}}(\omega)$ defined 
in ref.~\cite{MMS2}. This function is directly related with 
the contribution to $\chi(T)$ 
coming from the fluctuating clusters through eq.~(38) of ref.~\cite{MMS2}:
\begin{eqnarray}
\chi(T) = \xi^{-3} \int d\omega 
\frac{J_{{\rm res}}(\omega)}{\omega (\omega+T)} \, ,
\label{chires}
\end{eqnarray}
which is the analogue of  eq.~(5.15) of ref.~\cite{prb}. 
In fig.\ref{ires} we plot 
$\alpha_{C_2} \, J_{{\rm res}}(\omega)/J_{{\rm res}}(0)$ versus $\omega/\omega_0$ for $V_0=0.5$,
$c_{\gamma}=0.1$ and $\xi=20 \xi_0$ for various values of $C_2$ where
$\alpha_{C_2}$ is a scaling number that allows the maxima in each
curve to be at the same height ($\alpha_{0.5}=0.29$,$\alpha_1=1$,
$\alpha_{2}=3$, $\alpha_3 = 5.2$, $\alpha_4 = 7.4$, $\alpha_{10}=15.5$).
One can clearly see that the crossover frequency, which is essentially marked by
the maximum of each curve, occurs at low frequencies for $C_2>1$
and at large frequencies when $C_2<1$. This indicates that the region of
power law behavior shrinks, that is, as the
dissipation in the system increases, as $C_2$ decreases. The dissipation depends strongly on
the value of the coupling of the spins to the metallic environment, that is,
on the Kondo exchange coupling $J_K$ \cite{prb} and therefore it depends
on the microscopic details. These couplings change from system to
system leading to different values of $T^*$. It is very hard
to determine the value of these couplings from first principle calculations
and one has to rely on the experimental data in order to fit the value
of $C_2$ as one usually does in impurity systems in regards to the Kondo
temperature. This result explains why certain non-Fermi liquid materials
show strong power law behavior than others \cite{greg,marcio}.

We note that the conclusions of this work are relevant for
systems with Ising symmetry. Recent results for particular realizations
of Heisenberg quantum clusters \cite{millis_postdoc} and in studies of 
metallic Heisenberg magnets
\cite{vojta} have shown that tunneling is not suppressed and that
Griffiths-McCoy singularities are possible down to $T=0$. 
Nevertheless, essentially all systems of experimental interest where non-Fermi
liquid is observed have magnetic anisotropies due to spin-orbit
coupling and crystal field effects \cite{greg} and therefore are in
the Ising universality class discussed here. 

In summary, we have shown that the theory proposed by 
MMS in ref.~\cite{MMS2} has generically the same crossover behavior
that we predicted in ref.~\cite{prb} where for $T^*<T<\omega_0$
quantum Griffiths singularities with non-universal exponents,
$\chi(T) \propto T^{-1+\lambda}$, should be observed
and that for $T<T^*$ cluster freezing leads to $\chi(T) \propto
1/(T \ln(1/T))$ (see fig.\ref{phd}). 
The value of $T^*$, however, varies from
system to system and cannot be obtained within the theory. We
have argued in ref.~\cite{prb} $T^* \ll \omega_0$
and therefore power law behavior should be clearly visible in
a crossover region. This conclusion seems to be in agreement with
the experimental data in various materials \cite{greg,greg1,marcio}.

 {\it Acknowledgments}: We would like to acknowledge V.~Dobrosavljevic,
A.~J.~Millis, J.~Schmalian, D.~Morr, and T.~Vojta for 
helpful discussions.

\end{document}